\documentstyle [11pt]{article}
\def\ZzZ{{\hbox{\tenrm Z\kern-.31em{Z}}}}
\def\CcC{{\hbox{\tenrm C\kern-.45em{\vrule height.67em width0.08em depth-
.04em \hskip.45em }}}}

\def\mapbelow#1{\smash{\mathop{\longrightarrow}\limits_{#1}}}

\newcommand{\lab}{\label}
\newcommand{\non}{\nonumber}

\newcommand{\bc}{\begin{center}}
\newcommand{\ec}{\end{center}}
\newcommand{\be}{\begin{equation}}
\newcommand{\ee}{\end{equation}}
\newcommand{\bea}{\begin{eqnarray}}
\newcommand{\eea}{\end{eqnarray}}
\newcommand{\bs}{\begin{subequations}}
\newcommand{\es}{\end{subequations}}
\newcommand{\beq}{\begin{eqalignno}}
\newcommand{\eeq}{\end{eqalignno}}

\def\lab{\label}

\def\lf{\left}

\def\non{\nonumber}

\def\ri{\right}

\def\ka{\kappa}

\def\lab{\label}

\begin{document}

\bc {

{\bf Quantum noise, entanglement and chaos}

{\bf in the quantum field theory of mind/brain states}

\bigskip
\bigskip

Eliano Pessa$^\dag$ and Giuseppe Vitiello$^{\dag\dag}$

\medskip

$^{\dag}$Dipartimento di Psicologia, Universit\`a di Pavia, 27100
Pavia, Italy

$^{\dagger\dagger}$Dipartimento di Fisica ``E.R.Caianiello",
Universit\`a di Salerno, 84100 Salerno, Italy

INFN, Gruppo Collegato di Salerno and INFM, Sezione di Salerno

e-mail: pessa@unipv.it,  vitiello@sa.infn.it }
\bigskip

\ec

{\bf Abstract} We review the dissipative quantum model of brain
and present recent developments related with the r\^ole of
entanglement, quantum noise and chaos in the model.

\smallskip


\bigskip
\bigskip

\section{Introduction}
%

In this paper we present a short review of the dissipative quantum
model of brain (Vitiello, 1995; 2001)
and discuss some of its recent developments (Pessa and Vitiello,
2003)
related with the r\^ole of
entanglement, quantum noise and chaos in the model.

The quantum model of brain was originally formulated by Umezawa
and Ricciardi (Ricciardi and Umezawa, 1967)
and subsequently developed by Stuart, Takahashi and Umezawa
(Stuart, Takakhashi and Umezawa, 1978; 1979)
by Jibu and Yasue (Jibu and Yasue, 1995)
and by Jibu, Pribram and Yasue (Jibu, Pribram and Yasue, 1996)
. The formalism is the one of the quantum field theory (QFT). The
extension of the model to the dissipative dynamics has been worked
out more recently (Vitiello, 1995)
(see also (Alfinito and Vitiello, 2000; Pessa and Vitiello, 1999)
) and a general account in given in the book by G.V. {\it My
Double unveiled} (Vitiello, 2001)
.

The motivations at the basis of the formulation of the quantum
model of brain by Umezawa and Ricciardi trace back to the
laboratory observations leading, since the 1940's, Lashley to
remark that ``masses of excitations... within general fields of
activity, without regard to particular nerve cells''
(Lashley, 1942; Pribram, 1991) are involved in the determination
of behavior. In the middle of the 1960's Karl Pribram, also
motivated by experimental observations, started to formulate his
holographic hypothesis. Information appears indeed in such
observations to be spatially uniform ``in much the way that the
information density is uniform in a hologram''
(Freeman, 1990; 2000). While the activity of the single neuron is
experimentally observed in form of discrete and stochastic pulse
trains and point processes, the ``macroscopic'' activity of large
assembly of neurons appears to be spatially coherent and highly
structured in phase and amplitude
(Freeman, 1996; 2000).

Umezawa and Ricciardi, motivated thus by such an experimental
situation, formulated in 1967 (Ricciardi and Umezawa, 1967)
the quantum model of brain
as a many-body physics problem, namely by using the formalism
successfully tested in condensed matter experiments, the QFT with
spontaneous breakdown of symmetry. Such a formalism provides
indeed the only available theoretical tool capable to describe
long range correlations such the ones observed in the brain
presenting almost simultaneous responses in several regions to
some external stimuli. As a matter of fact, the understanding of
such correlations in terms of modern biochemical and
electrochemical processes is still lacking, which suggests that
these responses could not be explained in terms of single neuron
activity
 (Pribram 1971; 1991).

In QFT the dynamics (i.e. the Lagrangian) is in general invariant
under some group, say $G$, of continuous transformations.
Spontaneous breakdown of symmetry occurs
when the minimum energy state (the ground%
\index{vacuum state} state or vacuum) of the system is not
invariant under the full group $G$, but under one of its
subgroups. Then it can be shown
(Itzykson and Zuber, 1980; Umezawa, 1993)
that collective modes%
\index{collective mode}, the so--called Nambu--Goldstone%
\index{Nambu--Goldstone boson} (NG) boson modes, are dynamically
generated. Propagating over the whole system, these modes are the
carriers of the ordering information ({\it long range
correlation}): order manifests itself as a global property
dynamically generated. The long range correlation modes are
responsible for keeping the ordered pattern: they are coherently
{\it condensed} in the ground state.
In the crystal%
\index{crystal} case, for example, they keep the atoms trapped in
their lattice sites. The long range correlation thus forms a sort
of net, extending over all the system volume, which traps the
system components in the ordered pattern. This explain the
macroscopic collective behavior of the components as a ``whole''.

{\normalsize The observable specifying the ordered state%
\index{ordered state} is called the order parameter. It is a
measure of the
condensation%
\index{condensation} of the Nambu--Goldstone%
\index{Nambu--Goldstone boson} modes in the ground state and acts
as a macroscopic variable: it may be considered to be the {\it
code} specifying that specific ordered state. }

It is to be remarked that the spontaneous breakdown of symmetry is
possible since in QFT there exist infinitely many ground states or
vacua which are physically distinct (technically speaking, they
are ``unitarily inequivalent''). In quantum mechanics (QM), on the
contrary, all the vacua are physically equivalent and thus there
cannot be symmetry breakdown.

In the following section we will see how these features of QFT
apply to the brain model. The paper is organized as follows: in
Section 2 we shortly summarize the main aspects of the quantum
brain model. In Section 3 we consider its extension to the
dissipative dynamics and comment on the brain--environment
entanglement. In Section 4 and 5 we present the very recent
developments concerning quantum noise and chaos, respectively.
Section 6 is devoted to concluding remarks.

\medskip

\section{The quantum model of brain}

In this Section we present a short summary of the Ricciardi and
Umezawa model, closely following the ref: (Vitiello, 2001).

An essential, first requirement in the model is that stimuli
coming to the brain from the external world should be coded and
their effects on the brain should persist also after they have
ceased; this means that stimuli should be able to change the state
of the brain pre--existing the stimulation into another state
where the information is ``printed'' in a stable fashion. This
implies that the state where
information is recorded under the action of the stimuli must be a ground%
\index{vacuum state} state in order to realize the stability of
the recorded information; and that symmetry is broken
in order to allow the coding of the information. Recording%
\index{memory recording} of information is thus represented by
coherent condensation of NG bosons implied by the symmetry
breakdown. The NG collective modes are massless bosons, and thus
their condensation%
\index{condensation} in the vacuum%
\index{vacuum state} does not add energy to it: the stability of
the ordering, and therefore of the registered information, is thus
insured.

{\normalsize The order parameter%
\index{order parameter} is specific to the kind of symmetry of the
dynamics and its value is considered to be the {\it code}
specifying the information
printed in that ordered vacuum%
\index{vacuum state}.  Non--local properties, related to a code
specifying the system state, are dynamical features of quantum
origin: it is in this way that the stable and diffuse, non--local
character of memory is represented in the quantum model; it is
derived as a dynamical feature rather than as a property of
specific neural%
\index{neural networks} nets (which would be critically damaged by
local destructive actions). }

The mechanism of recall of the stored information is related to
the possibility of exciting collective modes%
\index{collective mode} out of the ground state. Suppose that an
ordered pattern was printed on the brain by
condensation%
\index{condensation} mechanism in the vacuum which was induced by
certain external stimuli. Though an order is stored, brain is not
conscious of this
because it is in the ground%
\index{vacuum state} state. However, when a similar external
stimulation comes in, it easily excites the massless boson
associated with the long
range correlation%
\index{long range correlation}. Since the boson is massless, any
small amount of energy can cause its excitation. During the time
of excitation, brain becomes conscious of the stored order
(memory). This explains recollection mechanism (Ricciardi and
Umezawa, 1967).
The excited modes have finite life--time and thus the recall
mechanism is a temporary activity of the brain, according indeed
to our common experience. This also suggests that the capability
to be ``alert'' or ``aware'' or to keep our ``attention'' focused
on certain subjects (information) for a short or a long time may
have to do with the short or long life--time of the excited modes
out of the brain ground state.

{\normalsize It may also happen that under the action of external
stimuli the brain may be put into an excited state, i.e. a
quasi--stationary state of
greater energy than the one of the ground%
\index{vacuum state} state. Such an excited state also carries
collective
modes%
\index{collective mode} in their non-minimum energy state. Thus
this state
also can support recording%
\index{memory recording} some information. However, due to its
higher energy such a state and the collective modes are not stable
and will sooner or
later decay: short--term memory is then modelled by the condensation%
\index{condensation} of long range correlation modes in the
excited states. Different types of short--term memory are
represented by different excitation levels in the brain state.

For a further analysis of the short--term memory mechanism in
terms of non--equilibrium phase transitions see also: (Sivakami
and Srinivasan, 1983).

The brain model should explain how memory remains stable and well
protected within a highly excited system, as indeed the brain is.
Such a ``stability" must be realized in spite of the permanent
electrochemical activity and the continual response to external
stimulation. The electrochemical activity must also, of course, be
coupled to the correlation modes which are triggered by external
stimuli. It is indeed the electrochemical activity observed by
neurophysiology that provides
(Stuart, Takahashi and Umezawa, 1978; 1979) a first response to
external stimuli.

This has suggested to model the memory mechanism as a separate
mechanism from the electrochemical processes of neuro-synapic
dynamics: the brain is then a ``mixed" system involving two
separate but interacting levels. The memory level is a quantum
dynamical level, the electrochemical activity is at a classical
level. The interaction between the two dynamical levels is
possible because of the specificity of the quantum dynamics: the
memory state is a {\it macroscopic quantum state} due to the
{\it coherence%
\index{coherence}} of the correlation modes.

{\normalsize The problem of the coupling between the quantum
dynamical level and the classical electrochemical level is then
reduced to the problem of the coupling of two macroscopic
entities. Such a coupling is analogous to the coupling
between classical acoustic waves and phonons%
\index{phonon} in crystals%
\index{crystal}. Acoustic waves are classical waves; phonons are
quantum NG long range modes. Nevertheless, their coupling is
possible since the macroscopic behavior of the crystal ``resides''
in the phonon modes, so that the coupling acoustic waves-phonon is
equivalently expressed as the coupling acoustic wave-crystal
(which is a perfectly acceptable coupling from a classical point
of view). }

We remark that the quantum variables in the quantum model of brain
are basic field variables (the electrical dipole field) and the
brain is described as a macroscopic quantum system. Stuart,
Takahashi and Umezawa (Stuart, Takakhashi and Umezawa, 1978)
have indeed remarked that ``it
is difficult to consider neurons as quantum objects''. In other
models of brain the relevant variables are binary variables
describing the neuron's on/off activity. However, in the quantum
model ``we do not intend", Ricciardi and Umezawa say ``to consider
necessarily the neurons as the fundamental units of the brain".

The quantum model of brain fits the
neurophysiological observations of memory nonlocality%
\index{nonlocality} and stability. However, several problems are
left open. One is that of memory capacity, the {\it overprinting%
\index{overprinting} problem}: Suppose a specific code
corresponding to a specific information
has been printed in the vacuum%
\index{vacuum state}. The brain then sets in that state and
successive
recording%
\index{memory recording} of a new, distinct (i.e. of different
code) information, under the action of a subsequent external
stimulus, is possible
only through a new condensation%
\index{condensation} process, corresponding to the new code. This
last condensation will superimpose itself on the former one ({\it
overprinting}), thus destroying the first registered information.

It has been shown (Vitiello, 1995)
that by taking into account the
dissipative character of brain dynamics may solve the problem of
memory capacity and quantum dissipation also turns out to be
crucial for the understanding of other functional features of the
brain. In the next Section we present a short summary of the
dissipative quantum model of brain.

\section{The dissipative quantum model of brain}

{\normalsize In the quantum model of brain the symmetry which
undergoes spontaneous breakdown under the action of the external
stimuli is the electrical dipole rotational symmetry. Water and
other biochemical molecules entering the brain activity are,
indeed, all characterized by a specific electrical dipole which
strongly constrains their chemical and physical behavior. Once the
dipole rotational symmetry has been broken (and information has
thus been recorded), then, as a consequence, time--reversal
symmetry is also
broken: {\it Before} the information recording%
\index{memory recording} process, the brain can in principle be in
anyone of the infinitely many (unitarily inequivalent) vacua. {\it
After} information has been recorded, the brain state is
completely determined and the brain cannot be brought to the state
configuration in which it was {\it before} the information
printing occurred. This is in fact the meaning of the well known
warning {\it ...NOW you know it!...}, which tells you that since
{\it now} you know, you are {\it another} person, not the same one
as {\it before}...Once you have known, you cannot go back in time.
}

Thus, ``getting information'' introduces {\it the arrow of time}
into brain dynamics; it introduces a partition in the time
evolution,  the {\it distinction} between the past and the future,
a distinction which did not exist {\it before} the information
recording. In other words, it introduce irreversibility, i.e.
dissipation. The brain is thus, unavoidably, an open system.

{\normalsize When the system under study is not an isolated system
it is customary in quantum theory to incorporate in the
treatment also the other systems (which constitute {\it %
the environment}) to which the original system is coupled. The
full set of systems then behaves as a single isolated (closed)
one. At the end of the required computations, one extracts the
information regarding the evolution of the original system by
neglecting the changes in the remaining systems. }

{\normalsize  In many cases, the specific details of the coupling
of our system with the environment may be very intricate and
changeable so that they are difficult to be measured and known.
One possible strategy is to average the effects of the coupling
and represent them, at some degree of accuracy, by means of some
``effective'' interaction. Another possibility is to take into
account the environmental
influence on the system by a suitable {\it choice} of the vacuum%
\index{vacuum state} state (the minimum energy state or ground
state).  The chosen vacuum thus carries the {\it signature} of the
reciprocal system--environment influence at a given time under
given boundary conditions. A change in the system--environment
reciprocal influence then would correspond to a change in the
choice of the system vacuum : the system ground state evolution or
``story'' is thus the story of the trade of the system with its
environment. The theory should then provide the equations
describing the system evolution ``through the vacua'', each vacuum
corresponding to the system ground state at each time of its
history.

{\normalsize In conclusion, in order to describe open quantum
systems first of all one needs to use QFT (Quantum Mechanics does
not have the many ''inequivalent'' vacua!). Then one also needs to
use the time variable as a label for the set of ground states of
the system
(Celeghini, Rasetti and Vitiello, 1990) : as the time (the label
value) changes, the
system moves to a ``new'' (physically inequivalent) ground%
\index{vacuum state} state (assuming continuous changes in the
boundary conditions determining the system--environment coupling).
Here, ``physically inequivalent'' means that the system
observables, such as the system energy, assume different values in
different inequivalent vacua, as is expected to happen in the case
of open systems. }

One thus gets a description for the open systems which is similar
to a collection of photograms: each photogram represents the
``picture" of our system at a given instant of time (a specific
time label value). Putting together these photograms in ``temporal
order" one gets a movie, i.e. the story (the evolution) of our
open system, which includes the system--environment interaction
effects.

The evolution of the ${\cal N}$--coded memory can be represented
as a trajectory of given initial condition running over
time--dependent states $|0(t)>_{{\cal N}}$, each one minimizing
the free energy functional. Recent results (Pessa and Vitiello,
2003; Vitiello, 2003)
show that such trajectories may have chaotic character. We will
discuss this in Section 5.

{\normalsize The mathematical representation of the environment
must explicitly satisfy the requirement that the energy lost by
the system must match the energy gained by the environment, and
viceversa. All other details of the system--environment
interaction may be taken into account
by the vacuum%
\index{vacuum state} structure of the system, in the sense above
explained. Then the environment may be represented in the simplest
way one likes, provided the energy flux balance is preserved. One
possible choice is to represent the environment as the
``time--reversed copy" of the system: time must be reversed since
the energy ``dissipated" by the system is ``gained" by
environment.
 }

{\normalsize In conclusion, the environment may be mathematically
represented as
the {\it time--reversed image%
\index{time--reversed image}} of the system, i.e. as the system
``double''. What the system loses, the environment gains, and
viceversa. }

{\normalsize Let $A_{\kappa}$ denotes the dipole wave quantum
(dwq) mode, namely the Nambu--Goldstone mode associated to the
spontaneous breakdown of rotational electrical dipole symmetry.
${\tilde A}_{\kappa}$ will denote its ``doubled mode''. The
$\tilde A$ mode is the ``time--reversed mirror image'' of the $A$
mode and represents the environment mode. Let ${\cal
N}_{A_{\kappa}}$ and ${\cal N}_{{ \tilde A}_{\kappa}}$ denote the
number of ${A_{\kappa}}$ modes and ${\tilde A }_{\kappa}$ modes,
respectively. The suffix $\kappa$ here generically denotes
kinematical variables ({\it e.g.} spatial momentum) or intrinsic
variables of the fields fully specifying the field degrees of
freedom.
 }

Notice that the ``tilde" or doubled mode is not just a
mathematical fiction. It corresponds to a real excitation mode
(quasiparticle) living in the system as an effect of its
interaction with the environment: the couples $A_{k}{\tilde
A}_{k}$ represent the correlation modes dynamically created in the
system as a response to the system--environment reciprocal
influence. It is the interaction between tilde and non--tilde
modes that controls the time evolution of the system: the
collective modes $A_{k}{\tilde A}_{k}$ are confined to live {\it
in} the system. They vanish as soon as the links between the
system and the environment are cut.

In the following Section we will see how these doubled modes may
be understood in terms of Wigner functions and how they are
related to quantum noise (Srivastava, Vitiello and Widom, 1995).

{\normalsize Taking into account dissipation requires
(Vitiello, 1995)
that the memory state, identified with the vacuum%
\index{vacuum state} ${|0>}_{{\cal N}}$ , is a condensate of {\it
equal
number} of $A_{\kappa }$ and ${%
\tilde{A}}_{\kappa }$ modes, for any $\kappa $ : such a
requirement ensures that the flow of the energy exchanged between
the system and the environment is balanced. Thus, the difference
between the number of tilde and non--tilde modes must
be zero: ${\cal N}_{A_{\kappa }}-{\cal N}_{{\tilde{A%
}}_{\kappa }}=0$, for any $\kappa $.

The label ${\cal N}$ in the vacuum symbol ${|0>}_{{\cal N}}$
specifies the set of integers $\{{\cal N}%
_{A_{\kappa }},~for~any~\kappa \}$ which indeed defines the
``initial value'' of the condensate, namely the {\it code}
associated to the information recorded at time $t_{0}=0$.

Note now that the requirement ${\cal %
N}_{A_{\kappa }}-{\cal N}_{{\tilde{A}}_{\kappa }}=0,$ for any
$\kappa $, does not uniquely fix the set $\{{\cal N}_{A_{\kappa
}},~{\rm for~any}~\kappa \}$.
Also ${|0>}_{{\cal N^{\prime }}}$
with ${\cal N^{\prime }}\equiv \{{\cal %
N^{\prime }}_{A_{\kappa }};
{\cal N^{\prime }}_{A_{\kappa }}-{\cal N^{\prime }%
}_{{\tilde{A}}_{\kappa }}=0,~for~any~\kappa \}$ ensures the energy
flow balance and therefore also ${|0>}_{{\cal N^{\prime }}}$ is an
available
memory state: it will correspond, however,
to a different code (i.e. $%
{\cal N^{\prime }}$) and therefore to a different information than
the one of code ${\cal N}$.

The conclusion is that fixing to zero the difference $%
{\cal N}_{A_{\kappa }}-{\cal N}_{{\tilde{A}}_{\kappa }}=0,$ for
any $\kappa $, leaves completely open the choice for the value of
the code ${\cal N}$.

Thus, infinitely many memory (vacuum) states, each one of them
corresponding to a different code ${\cal N}$, may exist: A huge
number of sequentially
recorded information data may {\it coexist} without destructive
interference%
\index{interference} since infinitely many vacua ${|0>}_{{\cal
N}}$, for all
${\cal N}$, are {\it independently} accessible in the sequential
recording%
\index{memory recording} process. Recording information
of code ${\cal %
N^{\prime }}$ does not necessarily produce destruction of
previously printed information of code ${\cal N}\neq {\cal
N^{\prime }}$, contrary to the non-dissipative case. In the
dissipative case the ``brain (ground) state'' may be represented
as the collection (or the superposition) of the full set of memory
states ${|0>}_{{\cal N}}$, for all ${\cal N}$. In the
non-dissipative case the ``${\cal N}$-freedom'' is missing and
consecutive information printing produces overprinting. }

The memory state is known (Vitiello, 1995)
to be a two-mode coherent state
(a generalized $SU(1,1)$ coherent state) and is given, at finite
volume $V$, by
\be\lab{M} {|0 \rangle}_{\cal N} =
\prod_k\;\frac{1}{\cosh\theta_{k}}\,\exp\left({-\tanh\theta_{k}
 A_k^{\dagger} {\tilde A}_{k}^{\dagger}}\right)\, |0\rangle_{0}
 ~, \ee
and, for all $\cal N$, ${}_{\cal N}\langle 0 | 0\rangle_{\cal N} =
1$.

$| 0\rangle_{\cal N}$ is an entangled state, namely it cannot be
factorized into two single--mode states. Indeed,
${|0\rangle}_{\cal N}$ can be written as
\be\label{M1}
  {|0\rangle}_{\cal N} = \left ( \prod_k\;\frac{1}{\cosh\theta_{k}} \right)\,\left (
|0 \rangle_{0} \otimes |{\tilde 0} \rangle_{0} - \sum_k
\;\tanh\theta_{k}
  \left( | A_k  \rangle \otimes |{\tilde A}_{k} \rangle
  \right)  + \dots \right)~,
\ee
where we have explicitly expressed the tensor product between the
tilde and non--tilde sector and dots stand for higher power terms.
Clearly, the second factor in the right hand side of the above
equation cannot be reduced to the product of two single--mode
components.

We remark that the entanglement is expressed by the unitary
inequivalence relation with the vacuum $|0 \rangle_{0} \equiv |0
\rangle_{0} \otimes |{\tilde 0} \rangle_{0}$:
\be\lab{ort2} {}_{\cal N}\langle 0 | 0 \rangle_{0} \mapbelow{V
\rightarrow \infty} 0 \quad \forall \, {\cal N} \neq {0} ~, \ee
which is verified only in the infinite volume limit. At finite
volume, a unitary transformation could disentangle the tilde and
non--tilde sectors: for a finite number of components their tensor
product would be different from the entangled state. However, this
is not the case in the infinite volume limit, where the summation
extends to an infinite number of components. In such a limit the
entanglement brain--environment is permanent. It cannot be washed
out: The entanglement mathematically represents the impossibility
to cut the links between the brain and the external world (a
closed, i.e. fully isolated, brain is indeed a dead brain
according to physiology).

Notice that memory states corresponding to different codes ${\cal
N} \neq {\cal N'}$,  ${|0\rangle}_{\cal N} $ and ${|0
\rangle}_{\cal N'}$, are each other unitarily inequivalent in the
infinite volume limit:
\be\lab{ort1} {}_{\cal N}\langle 0 | 0 \rangle_{\cal N'}
\mapbelow{V \rightarrow \infty} 0 \quad \forall \, {\cal N} \neq
{\cal N'} ~. \ee
{\normalsize This means that there does not exist in the infinite
volume
limit any unitary transformation which may transform one vacuum%
\index{vacuum state} of code ${\cal N}$ into another one of code ${\cal %
N^{\prime }}$: this fact, which is a typical feature of QFT,
guarantees that the corresponding printed information data are
indeed {\it different} or {\it distinguishable} ones (${\cal N}$
is a {\it good} code) and that each
information printing is also {\it protected} against interference%
\index{interference} from other information printing (absence of {\it %
confusion%
\index{confusion of memories}} among information data).

The average number ${\cal N}_{A_{\kappa}}$ is given by
\be\lab{num} {\cal N}_{A_{\kappa}} =
 {_{\cal N}}\langle  0| A_{\kappa}^{\dagger} A_{\kappa}{|0\rangle}
 _{\cal N} = \sinh^{2} \theta_{\kappa} ~,
\ee
and relates the $\cal N$--set, ${\cal N} \equiv \{ {\cal
N}_{A_{\kappa}} = {\cal N}_{{\tilde A}_{\kappa}}, \forall \kappa,
at~~t_{0}=0 \}$ to the $\theta$--set, $\theta \equiv \{
{\theta}_{\kappa}, \forall \kappa, at~~t_{0}=~0 \}$. We also use
the notation ${\cal N}_{A_{\kappa}}(\theta) \equiv {\cal
N}_{A_{\kappa}}$ and ${|0(\theta) \rangle} \equiv {|0
\rangle}_{\cal N}$. In general we may refer to ${\cal N}$ or,
alternatively and equivalently, to the corresponding $\theta$, or
viceversa.

The effect of finite (realistic) size of the system may spoil the
above mentioned unitary inequivalence. In the case of open
systems, in fact, transitions among (would be) unitary
inequivalent vacua may occur (phase transitions%
\index{phase transition}) for large but finite volume, due to
coupling with
the external environment. The inclusion of dissipation%
\index{dissipation} leads thus to a picture of the system ``living
over many
ground%
\index{vacuum state} states'' (continuously undergoing phase
transitions).
Note that even very weak (although above a certain threshold)
perturbations may drive the system through its macroscopic
configurations.
In this way, occasional (random) weak perturbations are recognized
to play an important r\^{o}le in the complex behavior of the brain
activity.

The possibility of transitions among differently coded vacua is a
feature of the model which is not completely negative: smoothing
out the exact unitary inequivalence among memory states has the
advantage of allowing the familiar
phenomenon of the ``association%
\index{association of memories}'' of memories: once transitions
among different memory states are ``slightly'' allowed, the
possibility of associations (``following a path of memories'')
becomes possible. Of course, these ``transitions'' should only be
allowed up to a certain degree in order
to avoid memory ``confusion%
\index{confusion of memories}'' and difficulties in the process of
storing ``distinct'' informational inputs
(Vitiello, 1995; Alfinito and Vitiello, 2000).
It is interesting to observe that Freeman, on the basis of
experimental observations, shows that noisy fluctuations at a
microscopic level may have a stabilizing effect on brain activity,
noise preventing to fall into some unwanted state (attractor) and
being an essential ingredient for the neural chaotic%
\index{chaos} perceptual apparatus (Freeman, 1990; 1996; 2000)
We will come back to consider quantum noise in Section 4.
 }

{\normalsize The dwq%
\index{dipole wave quanta} may acquire an effective non--zero mass
due to the effects of the system finite size (Vitiello, 1995;
Alfinito and Vitiello, 2000).
Such an effective mass will then act as a threshold for the
excitation energy of dwq so that, in order to trigger the recall
process, an energy supply equal or greater than such a threshold
is required. When the energy supply is lower than the required
threshold a ``difficulty in recalling'' may be experienced. At the
same time, however, the threshold may positively act as a
``protection'' against unwanted perturbations (including
thermalization%
\index{thermalization}) and contributes to the stability of the
memory state. In the case of zero threshold any replication signal
could excite the recalling and the brain would fall into a state
of ``continuous flow of memories''
(Vitiello, 1995). }

Summarizing,  the brain system may be viewed as a complex system
with (infinitely) many macroscopic configurations (the memory
states). Dissipation%
\index{dissipation}, which is intrinsic to the brain dynamics, is
recognized
to be the root of such a complexity, namely of the huge memory capacity%
\index{memory capacity}.

Of course, the brain has several structural and dynamical levels
(the basic level of coherent condensation%
\index{condensation} of dwq%
\index{dipole wave quanta}, the cellular cytoskeleton%
\index{cytoskeleton} level, the neuronal dendritic level, and so
on) which coexist, interact among themselves and influence each
other's functioning. Dissipation introduces the further richness
of the replicas or degenerate vacua at the basic quantum level.
The crucial point is that the different levels of organization are
not simply structural features of the brain, their reciprocal
interaction and their evolution is intrinsically related to the
basic quantum dissipative dynamics.

The brain's functional
stability is ensured by the system's ``coherent response%
\index{coherent response}'' to the multiplicity of external
stimuli. Thus dissipation also seems to suggest a solution to the
so called {\it binding problem}, namely the understanding of the
unitary response and behavior of apparently separated units and
physiological structures of the brain.

When considering dwq with time--dependent frequency, modes with
longer life--time are found to be the ones with higher momentum.
Since the momentum is proportional to the reciprocal of the
distance over which the mode can propagate, this means that modes
with shorter range of propagation will survive longer. On the
contrary, modes with longer range of propagation will decay
sooner. The scenario becomes then particularly interesting since
this mechanism may
produce the formation%
\index{domain formation} of ordered domains of finite different
sizes with different degree of stability: smaller domains would be
the more stable ones. Remember now that the regions over which the
dwq propagate are the domains where ordering (i.e. symmetry
breakdown) is produced. Thus we arrive
at the dynamic formation%
\index{domain formation} of a hierarchy (according to their
life--time or equivalently to their sizes) of ordered domains
(Alfinito and Vitiello, 2000).

\medskip

\section{Quantum noise}

In this Section, by resorting to previous results on dissipative
quantum systems (Srivastava, Vitiello and Widom, 1995; Blasone et
al., 1998),
we show that the doubled variables account for the quantum noise
effects in the fluctuating random force in the system--environment
coupling. This opens a new perspective in the quantum model of
brain, and fits well with some experimental observations in the
brain behavior (Freeman, 1990; 1996; 2000).

Our discussion will also show how the doubling of the degrees of
freedom is related to the Wigner function and to the density
matrix formalism.

Before considering dissipation, let us consider the case of zero
mechanical resistance. The Hamiltonian for an isolated particle is
\be\lab{H} H=- \frac{\hbar^2}{2m}\left(\frac{\partial}{\partial
x}\right)^2 +V(x)~, \ee
and the expression for the Wigner function is (Feynman, 1972;
Haken, 1984)
%
\be\lab{W} W(p,x,t) = \frac{1}{2\pi \hbar}\int {\psi^* \lf(x -
\frac{1}{2}y,t\ri)\psi \lf(x + \frac{1}{2}y,t\ri)
e^{\lf(-i\frac{py}{\hbar}\ri)}dy} ~. \ee
The associated density matrix function is
\be\lab{8} W(x,y,t)=\langle x+\frac{1}{2}y|\rho
(t)|x-\frac{1}{2}y\rangle = \psi^* \lf(x-\frac{1}{2}y,t\ri)\psi
\lf(x+\frac{1}{2}y,t\ri)~, \ee
with equation of motion given by
\be\lab{5} i\hbar \frac{d \rho}{dt}=[H,\rho ]~. \ee
By introducing the notation
\be\lab{7} x_{\pm}=x\pm \frac{1}{2}y~, \ee
Eq. (\ref{5}) is written in the coordinate representation as
\bea \non && i\hbar \frac{\partial}{\partial t}\langle x_+|\rho
(t)|x_-\rangle=
\\ \lab{6}
&&\lf\{ -\frac{\hbar^2}{2m}\lf[\lf(\frac{\partial}{\partial
x_+}\ri)^2-\lf(\frac{\partial}{\partial x_-}\ri)^2\ri]
+[V(x_+)-V(x_-)] \ri\}\langle x_+|\rho (t)|x_-\rangle , \eea
namely, in terms of  $x$ and $y$, we have
\be\lab{9a} i\hbar \frac{\partial }{\partial t} W(x,y,t)={\cal
H}_o W(x,y,t) ~, \ee
\be\lab{9b} {\cal H}_o=\frac{1}{m}p_xp_y
+V\lf(x+\frac{1}{2}y\ri)-V\lf(x-\frac{1}{2}y\ri), \ee
with $p_x=-i\hbar\frac{\partial}{\partial x}, \ \
p_y=-i\hbar\frac{\partial}{\partial y}$. The Hamiltonian
(\ref{9b}) may be constructed from the Lagrangian
\be\lab{10}
 {\cal L}_o=m
\dot{x}\dot{y}-V\lf(x+\frac{1}{2}y\ri)+V\lf(x-\frac{1}{2}y\ri).
\ee
We thus see that the density matrix and the Wigner function
formalism requires the introduction of a ``doubled" set of
coordinates, $x_{\pm}$, or, alternatively, $x$ and $y$.

In the case of the particle interacting with a thermal bath at
temperature $T$, the interaction Hamiltonian between the bath and
the particle is taken as
\be\lab{11} H_{int}=-fx, \ee
where $f$ is the random force on the particle at the position $x$
due to the bath.

In the Feynman-Vernon formalism, it can be shown (Srivastava,
Vitiello and Widom, 1995)
that the effective action for the particle has the form
\be\lab{12} {\cal A}[x,y]=\int_{t_i}^{t_f}dt\,{\cal
L}_o(\dot{x},\dot{y},x,y) +{\cal I}[x,y], \ee
where ${\cal L}_o$ is defined in Eq.(\ref{10}) and
\bea\nonumber {\cal
I}[x,y]&=&\frac{1}{2}\int_{t_i}^{t_f}dt\,[x(t)F_y^{ret}(t)+
y(t)F_x^{adv}(t)]
\\ \lab{29}
&&+\frac{i}{2\hbar}\int_{t_i}^{t_f}\int_{t_i}^{t_f}dt ds
\,N(t-s)y(t)y(s) ~, \eea
where the retarded force on $y$ and the advanced force on $x$ are
given in terms of the retarded and advanced Greens functions and
$N(t-s)$ denotes the quantum noise in the fluctuating random force
given by
\be\lab{21} N(t-s)=\frac{1}{2}\langle f(t)f(s)+f(s)f(t)\rangle ~.
\ee
The symbol $\langle...\rangle$ denotes average with respect to
thermal bath. One can then show (see (Srivastava, Vitiello and
Widom, 1995))
for the details of the derivation) that the real and the imaginary
part of the action are given by
\be\lab{30a} {\cal R}e{\cal A}[x,y]=\int_{t_i}^{t_f}dt\,{\cal L},
\ee
\be\lab{30b} {\cal L}\,=\,m \dot{x}\dot{y}-\lf[V(x+\frac{1}{2}
y)-V(x-\frac{1}{2}y)\ri] + \frac{1}{2}\lf[x F_y^{ret} + y
F_x^{adv}\ri], \ee
and
\be\lab{30c} {\cal I}m{\cal A}[x,y]=
\frac{1}{2\hbar}\int_{t_i}^{t_f}\int_{t_i}^{t_f}dt ds
\,N(t-s)y(t)y(s). \ee
respectively. These results, Eqs.(\ref{30a}),~(\ref{30b}), and
(\ref{30c}), are rigorously exact for linear passive damping due
to the bath. They show that, in the classical limit ``$\hbar
\rightarrow 0$'', nonzero $y$ yields an ``unlikely process'' due
to the large imaginary part of the action implicit in
Eq.(\ref{30c}). On the contrary, at quantum level nonzero $y$
accounts for quantum noise effects in the fluctuating random force
in the system--environment coupling arising from the imaginary
part of the action} (Srivastava, Vitiello and Widom, 1995).

In the case one considers the approximation to Eq.(\ref{30b}) with
$F_y^{ret}=\gamma\dot{y}$ and $F_x^{adv}=-\gamma\dot{x}$, and puts
$V\lf(x \pm \frac{1}{2}y\ri) = \frac{1}{2}\ka (x \pm
\frac{1}{2}y)^{2}$, then the damped harmonic oscillator (dho) for
the $x$ variable and the complementary equation for the $y$
coordinate can be derived
\bea\lab{3cx} m \ddot{x}+\gamma\dot{x}+\ka x&=&0,
\\ \lab{3cy} m \ddot{y}-\gamma\dot{y}+\ka y&=&0 . \eea
The $y$-oscillator is thus recognized to be the time--reversed
image of the $x$-oscillator. Of course, from the manifold of
solutions to Eqs.(\ref{3cx}),~(\ref{3cy}) we could choose the ones
for which the $y$ coordinate is constrained to be zero. Then we
obtain the classical damped oscillator equation from a lagrangian
theory at the expense of introducing an ``extra'' coordinate $y$,
later constrained to vanish.

It should be stressed, however, that {\it the r\^ole of the
``doubled" $y$ coordinate is absolutely crucial in the quantum
regime since there it accounts for the quantum noise in the
fluctuating random force in the system--environment coupling}, as
shown above. Reverting, instead, from the classical level to the
quantum level, the loss of information occurring  at the classical
level due to dissipation manifests itself in terms of ``quantum"
noise effects arising from the imaginary part of the action, to
which the $y$ contribution is indeed crucial.

Going back to the dissipative quantum model of brain, we note that
the classical equations for the dho $x$ and its time--reversal
image $y$, Eqs. (\ref{3cx}) and (\ref{3cy}), are associated in the
canonical quantization procedure to the quantum operators  $A$ and
$\tilde A$ (Vitiello, 1995),
respectively. When we consider the quantum field theory, the $A$
and the $\tilde A$ operators get labelled by the (continuously
varying) suffix $\kappa$ and for each $\kappa$ value we have a
couple of equations of the type (\ref{3cx}) and (\ref{3cy}) for
the field amplitudes (Celeghini, Rasetti and Vitiello, 1992).

In conclusion, we have seen that the doubling of the degrees of
freedom discussed in the previous sections accounts for the
quantum noise in the fluctuating random force coupling the system
with the environment (the bath). On the other hand, we have also
recognized the entanglement between the tilde and the non--tilde
modes. Thus we conclude that brain processes are intrinsically and
inextricably dependent on the quantum noise in the fluctuating
random force in the brain-environment coupling. It is interesting
to mention, in this respect, the r\^ole of noise in neurodynamics
which has been noticed by Freeman in his laboratory observations
(Freeman, 1990; 1996; 2000).

In the following section we discuss another feature built--in in
the dissipative quantum model: the chaotic behavior of the
trajectories in the space of the memory states.

\medskip

\section{Chaos and memory states}

In this section we analyze the time evolution of the memory
states.

We denote by ${|0 (t) \rangle}_{\cal N}$ the memory state at time
$t$ and refer to the space of the memory states, for all $\cal N$
and at any time $t$, as to the ``memory space". In this space the
memory state ${|0 (t) \rangle}_{\cal N}$ may be thought as a
``point" labelled by a given $\cal N$--set (or $\theta$--set) and
by a given value of $t$ and points corresponding to different
$\cal N$ (or $\theta$) sets and different $t$'s are distinct
points (do not overlap, cf. Eq. (\ref{ort1}) and Eq. (\ref{tt})
below). As mentioned, the memory states can be understood as the
vacuum states of corresponding Hilbert spaces. In QFT these
Hilbert spaces are denoted as the representations of the canonical
commutation relations for the operators $A$ and $\tilde A$, which
in the infinite volume limit are unitarily inequivalent. So, the
memory space may also be denoted as the ``space of the (unitarily
inequivalent) representations'', each representation being
represented by a ``point'' in such a larger memory space.

We show that trajectories over the memory space (the
representation space) may be chaotic trajectories.

The requirements for chaotic behavior in non--linear dynamics can
be formulated as follows (Hilborn, 1994):

i)~ the trajectories are bounded and each trajectory does not
intersect itself (trajectories are not periodic).


ii)~~there are no intersections between trajectories specified by
different initial conditions.



iii) trajectories of different initial conditions are diverging
trajectories.


At finite volume $V$, the memory state ${|0 (t) \rangle}_{\cal
N}$, to which the memory state, say at $t_{0}=0$, $| 0
\rangle_{\cal N}$ evolves, is given (Vitiello, 1995) by
%
%
\be\lab{timev} | 0(t) \rangle_{\cal N} = \prod_{\kappa}
{1\over{\cosh{(\Gamma_{\kappa} t
  - {\theta}_{\kappa} )}}} \exp{
\left ( \tanh {(\Gamma_{\kappa} t - {\theta}_{\kappa}  )}
A^{\dag}_{\kappa}{\tilde A}^{\dag}_{\kappa} \right )}
|0\rangle_{0} ~, \ee
which also is an entangled, $SU(1,1)$ generalized coherent state.
$\Gamma_{\kappa}$ is the damping constant implied by the
dissipation (Vitiello, 1995).
Note that for any $t$
\be\lab{nr} {}_{\cal N}\langle 0(t) | 0(t)\rangle_{\cal N}  = 1 ~.
\ee

In the infinite volume limit we have (for $ {\int \! d^{3} \kappa
\, \Gamma_{\kappa}}$ finite and positive)
\be\lab{t} {}_{\cal N}\langle 0(t) | 0\rangle_{\cal N}
\mapbelow{V \rightarrow \infty} 0 \quad ~~ \forall \,
 t \quad ,
\ee
\be\lab{tt} {}_{\cal N}\langle 0(t) | 0(t') \rangle_{\cal N}
\mapbelow{V \rightarrow \infty} 0 \quad ~~ \forall \, t\, , t'
\quad , \quad t \neq t' \quad . \ee
States $|0(t)\rangle_{\cal N} $ (and the associated Hilbert spaces
$\{| 0(t) \rangle_{\cal N} \}$) at different time values $t \neq
t'$ are thus unitarily inequivalent in the infinite volume limit.

Time evolution of the memory state $|0\rangle_{\cal N}$ is thus
represented as the (continuous) transition through the
representations $\{| 0(t) \rangle_{\cal N} \}$ at different $t$'s
(same $\cal N$) , namely by the ``trajectory" through the
``points" $\{| 0(t) \rangle_{\cal N} \}$ in the space of the
representations. The trajectory "initial condition" at $t_{0} = 0$
is specified by the $\cal N$--set.

It is known (Manka, Kuczynski and Vitiello, 1986; Del Giudice et
al., 1988)
(see also (Vitiello, 2003)
for a recent discussion) that trajectories of this kind are
classical trajectories: transition from one representation to
another inequivalent one would be strictly forbidden in a quantum
dynamics.

We now observe that the trajectories are {\it bounded} in the
sense of Eq. (\ref{nr}), which shows that the ``length"  of the
``position vectors" (the state vectors at time $t$) in the
representation space is finite (and equal to one) for each $t$ (by
resorting to the properties of the $SU(1,1)$ group, one can show
that the set of points representing the coherent states
$|0(t)\rangle_{\cal N}$ for any $t$ is isomorphic to the union of
unit circles of radius ${r_{\kappa}}^{2} =
\tanh^{2}(\Gamma_{\kappa}t - \theta_{\kappa})$ for any $\kappa$
(Perelomov 1986; Pessa and Vitiello, 2003).

We also note that Eqs. (\ref{t}) and (\ref{tt}) express the fact
that the trajectory does not crosses itself as time evolves (it is
not a periodic trajectory): the ``points" $|0(t)\rangle_{\cal N}$
and $|0(t')\rangle_{\cal N}$ through which the trajectory goes,
for any $t$ and $t'$, with $t \neq t'$, after the initial time
$t_{0} = 0$, never coincide. The requirement $i)$ is thus
satisfied.

In the infinite volume limit, Eqs. (\ref{t}) and (\ref{tt}) also
hold for ${\cal N} \neq {\cal N'}$, i.e. we also have
\be\lab{tn} {}_{\cal N}\langle 0(t) | 0\rangle_{\cal N'}
\mapbelow{V \rightarrow \infty} 0 \quad ~~ \forall \,
 t \quad , ~~\forall \, {\cal N} \neq {\cal N'}
\ee
\be\lab{ttn} {}_{\cal N}\langle 0(t) | 0(t') \rangle_{\cal N'}
\mapbelow{V \rightarrow \infty} 0 \quad ~~ \forall \, t\, , t'
\quad , ~~\forall \, {\cal N} \neq {\cal N'} ~. \ee
The derivation of Eqs. (\ref{tn}) and (\ref{ttn}) rests on the
fact that in the continuum limit, for given $t$ and $t'$ and for
${\cal N} \neq {\cal N'}$, $\cosh({\Gamma}_{\kappa}t -
{\theta}_{\kappa} + {\theta'}_{\kappa})$ and
$\cosh({\Gamma}_{\kappa}(t - t') - {\theta}_{\kappa} +
{\theta'}_{\kappa})$, respectively, are never identically equal to
one {\it for all} $\kappa$.

Notice that Eq. (\ref{ttn}) is true also for $t = t'$ for any
${\cal N} \neq {\cal N'}$. Eqs. (\ref{tn}) and (\ref{ttn}) thus
tell us that trajectories specified by different initial
conditions (${\cal N} \neq {\cal N'}$) never cross each other.
Thus, also requirement ii) is satisfied.

We remark that, in the infinite volume limit, due to the property
ii) no {\it confusion} (interference) arises among different
memories, even as time evolves. In realistic situations of finite
size coherent domains, some {\it association} of memories may
become possible. In such a case, this means the at a ``crossing"
point between two, or more than two, trajectories, from one of
these trajectories one can switch to another one which there
crosses. This may be felt indeed as association of memories or as
``switching" from one information to another one.

The average number of modes of type $A_{\kappa}$ is given, at each
instant $t$, by
\be\lab{Nt} {\cal N}_{A_{\kappa}}(\theta,t) \equiv {}_{\cal
N}\langle 0(t) | A_{\kappa}^{\dagger} A_{\kappa} | 0(t)
\rangle_{\cal N}  = \sinh^{2}\bigl ( \Gamma_{\kappa} t -
{\theta}_{\kappa} \bigr ) ~, \ee
and similarly for the modes of type ${\tilde A}_{\kappa}$. This
number can been shown to satisfy the Bose distribution and thus it
is actually a statistical average (Vitiello, 1995).
From Eq. (\ref{Nt}) we see that at a time $t = \tau$, with $\tau$
the largest of the values ${t_{\kappa}} \equiv {
{{\theta}_{\kappa}}\over{\Gamma_{\kappa}}}$, the memory state
$|0\rangle_{\cal N}$ is reduced (decayed) to the "empty" vacuum
$|0\rangle_{0}$: the information has been {\it forgotten}, the
${\cal N}$ code is decayed. The time $t = \tau$ can be taken to be
the life--time of the memory of code $\cal N$ (for details on this
point we refer to (Alfinito and Vitiello, 2000),
where a detailed analysis of the
life--time of the $\kappa$--modes has been made). We conclude that
the time evolution of the memory state leads to the "empty" vacuum
$|0\rangle_{0}$ which acts as a sort of attractor state. However,
as time goes on, i.e. as $t$ gets larger than $\tau$, we have
\be\lab{tt0} \lim_{t\to \infty} {}_{\cal N}\langle 0(t) |
0\rangle_{0} \, \propto \lim_{t\to \infty}
 \exp{\left ( -t  \sum_{\kappa}  \Gamma_{\kappa}  \right )} = 0 ~,
\ee
which tells us that the state $|0(t)\rangle_{\cal N}$ "diverges"
away from the attractor state $| 0\rangle_{0}$ with exponential
law (we are always assuming ${\sum_{\kappa} \Gamma_{\kappa} >
0}$).

It is interesting to observe that in order to avoid to fall into
such an attractor, i.e. in order to not forget certain
information, one needs to "restore" the ${\cal N}$ code by
"refreshing" the memory by {\it brushing up} the subject (external
stimuli maintained memory). This means that one has to recover the
whole ${\cal N}$--set (if the whole code is ``corrupted"), or
``pieces" of the memory associated to those ${\cal N}_{\kappa}$,
for certain $\kappa$'s, which have been lost at $t_{\kappa} = {
{{\theta}_{\kappa}}\over{\Gamma_{\kappa}}}$. The operation of
restoring the code appears to be a sort of ``updating the
register" of the memories since it amounts to reset the memory
code (and clock) to the (updated) initial time $t_{0}$. We also
observe that even after the time $\tau$ is passed by, the code
$\cal N$ may be recovered provided $t$ is not much larger than
$\tau$ (namely, as far as the approximation of
$\cosh({\Gamma}_{\kappa}t - {\theta}_{\kappa}) \approx \exp{\left
( -t \sum_{\kappa} \Gamma_{\kappa}  \right )}$ does not hold, cf.
Eq. (\ref{tt0})).

We now consider the variation in time of the ``distance'' between
trajectories in the memory space, i.e. the variation in time of
the difference between two different codes, ${\cal N} \neq {\cal
N'}$ ($\theta \neq {\theta}'$), corresponding to different initial
conditions for two trajectories. At time $t$, each component
${\cal N}_{A_{\kappa}}(t)$ of the code ${\cal N} \equiv \{ {\cal
N}_{A_{\kappa}} = {\cal N}_{{\tilde A}_{\kappa}}, \forall \kappa,
at~~ t_{0}=0 \}$ is given by the expectation value in the memory
state of number operator $A_{\kappa}^{\dagger} A_{\kappa}$. The
difference is then (cf. Eq. (\ref{Nt}):
$$
\Delta {\cal N}_{A_{\kappa}}(t) \equiv {\cal
N'}_{A_{\kappa}}(\theta',t) - {\cal N}_{A_{\kappa}}(\theta,t) =
$$
\be\lab{co1} = \sinh^{2}\bigl ( \Gamma_{\kappa} t  -
\theta_{\kappa} + {\delta \theta_{\kappa}} \bigr ) -
\sinh^{2}\bigl ( \Gamma_{\kappa} t -  \theta_{\kappa} \bigr )
\approx \sinh \bigl ( 2 (\Gamma_{\kappa} t - \theta_{\kappa} )
\bigr ){\delta \theta_{\kappa}} ~, \ee
where ${\delta \theta_{\kappa}} \equiv {\theta}_{\kappa} -
{\theta'}_{\kappa}$ (which without loss of generality may be
assumed to be greater than zero), and the last equality holds for
small ${\delta \theta_{\kappa}}$ (i.e. for a very small difference
in the initial conditions of the two memory states). The
time--derivative then gives
\be\lab{co3} \frac{\partial}{\partial t}\Delta {\cal
N}_{A_{\kappa}}(t) = 2 {\Gamma_{\kappa}} \cosh \bigl ( 2
(\Gamma_{\kappa} t -  \theta_{\kappa}) \bigr ){\delta
\theta_{\kappa}} ~, \ee
which shows that the difference between ${\cal N}_{A_{\kappa}}$'s,
originally even slightly different, is a growing function of time.
For enough large $t$ the modulus of the difference $\Delta {\cal
N}_{A_{\kappa}}(t)$ and its variation in time diverge as
$\exp{(2\Gamma_{\kappa} t)}$, for all $\kappa$'s. For each
$\kappa$, $2\Gamma_{\kappa}$ play thus the r\^ole similar to the
one of the Lyapunov exponent.

Thus we conclude that trajectories in the memory space differing
by a small variation ${\delta \theta}$ in the initial conditions
are diverging trajectories as time evolves.

It should be remarked that, as shown by Eq. (\ref{co1}), the
difference between specific $\kappa$--components of the codes
$\cal N$ and $\cal N'$ may become zero at a given time $t_{\kappa}
= \frac{\theta_{\kappa}}{\Gamma_{\kappa}}$. However, this does not
mean that the difference between the codes $\cal N$ and $\cal N'$
becomes zero. They are made, indeed, by a large number (infinite
number, in the continuum limit) of components and they are still
different codes even if a finite number of their components are
equal. On the contrary, always for very small ${\delta
\theta_{\kappa}} \equiv {\theta}_{\kappa} - {\theta'}_{\kappa}$,
suppose that the time interval $\Delta t = \tau_{max}
-\tau_{min}$, with $\tau_{min}$ and $\tau_{max}$ the smallest and
the largest, respectively, of the $t_{\kappa} =
\frac{\theta_{\kappa}}{\Gamma_{\kappa}}$, for {\it all}
$\kappa$'s, is ``very small'', then in such a $\Delta t$ the codes
are ``recognized'' to be ``almost'' equal. In such a case, Eq.
(\ref{co1}) expresses the ``recognition'' (or recall) process and
we see how it is possible that ``slightly different'' ${\cal
N}_{A_{\kappa}}$--patterns (or codes) are ``identified''
(recognized to be the ``same code'' even if corresponding to
slightly different inputs). Roughly, $\Delta t$ may be taken as a
measure of the ``recognition time''.

We finally recall that $\sum_{\kappa} E_{\kappa} {\dot {\cal
N}}_{A_{\kappa}} dt = \frac{1}{\beta} dS_{A}$ (see (Vitiello,
1995)),
where $E_{\kappa}$ is the energy of the mode $A_{\kappa}$, $\beta
= \frac{1}{k_{B}T}$, $k_{B}$ the Boltzmann constant, $dS_{A}$ is
the entropy variation associated to the modes $A$ and ${\dot {\cal
N}}_{A_{\kappa}}$ denotes the time derivative of ${\cal
N}_{A_{\kappa}}$. Eq. (\ref{co3}) then leads to the relation
between the differences in the variations of the entropy and the
divergence of trajectories of different initial conditions:
\be\lab{co4} \Delta \sum_{\kappa} E_{\kappa}{\dot {\cal
N}}_{A_{\kappa}}(t) dt = \sum_{\kappa} 2 E_{\kappa}
{\Gamma_{\kappa}}^{2} \cosh \bigl ( 2 (\Gamma_{\kappa} t  -
\theta_{\kappa}) \bigr ){\delta \theta_{\kappa}} dt =
\frac{1}{\beta} \bigl ( dS'_{A} - dS_{A} \bigr ) ~. \ee

In conclusion, also the requirement iii) is satisfied.

We thus conclude that the trajectories in the memory space may
exhibit chaotic behavior.  This is a feature which fits
experimental observations by Freeman (Freeman, 1990; 1996; 2000)
who indeed finds characteristic chaotic behavior in neural
aggregates of the olfactory system of laboratory pets.

\medskip

\section{Concluding remarks}

We have presented a short review of the dissipative quantum model
of brain and we have shown that the doubling of the system degrees
of freedom account for the quantum noise in the fluctuating random
force coupling the system with the environment. Due to the
permanent entanglement brain--environment, quantum noisy effects
are intrinsically present in the brain dynamics. Moreover, we have
seen that time evolution in the memory space may present chaotic
behavior. It is interesting to remark that laboratory observations
show an important r\^ole in brain dynamics of noise and of chaos
(Freeman, 1990; 1996; 2000).
In the dissipative model noise and chaos turn out to be natural
ingredients of the model. In particular the chaotic behavior of
the trajectories in memory space may account of the high
perceptive resolution in the recognition of the inputs. Indeed,
small differences in the codes associated to external inputs may
lead to diverging differences in the corresponding memory paths.
On the other side, codes ``almost'' equal in all of their
components may easily be recognized as being the ``same'' code
(code identification, or ``code--pattern'' recognition).

Further work on these subjects is in progress (Pessa and Vitiello,
2003).

\bigskip
\bigskip

{\bf Acknowledgments}
\smallskip

 We acknowledge MIUR, INFN and INFM for partial financial support.
 One of the authors (G.V.) is much grateful to Harald Atmanspacher
 for inviting him at the International Workshop on Aspects of
 Mind--Matter Research, June 2003, Wildbad Kreuth (Germany), where
 part of the results of the present paper were reported.

\end{document}